\documentclass[prb,twocolumn,showpacs,amsmath,amssymb]{revtex4}

\usepackage{graphicx}
\usepackage{dcolumn}
\usepackage{bm}
\usepackage{gensymb}
\usepackage{tabularx}

\begin{document}

\title{Type I Superconductivity in YbSb$_2$ Single Crystals}

\author{Liang L. Zhao,$^{1}$ Stefan Lausberg,$^{2}$ H. Kim,$^{3}$ M.~A. Tanatar,$^{3}$ Manuel Brando,$^{2}$ R. Prozorov,$^{3}$ and E. Morosan$^{1}$}
\affiliation{$^1$Department of Physics \& Astronomy, Rice University, Houston, TX 77005, USA\\
             $^2$Max Planck Institute for Chemical Physics of Solids, 01187 Dresden, Germany\\
             $^3$Ames Laboratory and Department of Physics \& Astronomy, Iowa State University, IA 50011, USA}

\date{\today}

\begin{abstract}
We present evidence of type I superconductivity in YbSb$_2$ single crystals, from DC and AC magnetization, heat capacity and resistivity measurements. The critical temperature and critical field are determined to be $T_c\approx$ 1.3 K and $H_c\approx$ 55 Oe. A small Ginzburg-Landau parameter $\kappa$ = 0.05, together with typical magnetization isotherms of type I superconductors, small critical field values, a strong Differential Paramagnetic Effect (DPE) signal, and a field-induced change from second to first order phase transition, confirm the type I nature of the superconductivity in YbSb$_2$. A possible second superconducting state is observed in the radiofrequency (RF) susceptibility measurements, with $T_{c}^{(2)}\approx$ 0.41 K and $H_{c}^{(2)}\approx$ 430 Oe.

\end{abstract}


\maketitle

\section{Introduction}

A long held empirical belief has been that type I superconductors are generally elementary metals and metalloids, while the majority of superconducting compounds exhibit type II behavior. Among the vast array of known binary and ternary superconductors, the number of systems with type I superconductivity is notably limited. \cite{CxK,RPd2Si2andRRh2Si2,TaSi2,Ag5Pb2O6,LaRhSi3,RGa3,YamaguchiYbSb2}

YbSb$_2$ was first claimed to be a type I superconductor by Yamaguchi \emph{et al.}\cite{YamaguchiYbSb2}, solely based on the shape of one $M(H)$ isotherm at 0.4 K. Subsequently, a limited number of studies of the physical properties of YbSb$_2$ have been published. Among those, Sato \emph{et al.} reported results of density functional theory (DFT) calculations, resistivity and de Haas-van Alphen measurements, which revealed a quasi-two-dimensional Fermi surface.\cite{SatodHvA} Two other brief reports of resistivity under pressure \cite{pressureYbSb2} and NQR measurements\cite{SbNQR} indicated that $T_c$ is suppressed under pressure $p=$ 0.4 GPa, and that YbSb$_2$ is likely a weakly coupled $s$-wave superconductor. Given the scarcity of type I superconducting compounds, and the lack of a thorough characterization of the magnetic and thermodynamic properties of YbSb$_2$, a detailed analysis of the superconducting ground state in this compound is needed. In this paper, we report results of DC and AC magnetization, heat capacity, resistivity and magnetic penetration depth, confirming the superconducting ground state with $T_c\approx$ 1.3 K and $H_c\approx$ 55 Oe. A discussion of the superconducting parameters, based on the BCS and Ginzburg-Landau (GL) theories, is also provided. The shape of the $M(H)$ curves and a second to first order phase transition in specific heat below $T_c$, together with large differential paramagnetic effect (DPE), small critical field $H_c$ and GL parameter $\kappa\ll1/\sqrt{2}$ provide strong evidence for the type I superconductivity in YbSb$_2$. Moreover, the RF susceptibility data reveal a possible second superconducting transition with $T_{c}^{(2)}\approx$ 0.41 K and $H_{c}^{(2)}\approx$ 430 Oe.

\section{Experimental Methods}
YbSb$_2$ single crystals were synthesized by flux growth technique, using excess amount of Sb. Elemental Yb (Ames Lab, 99.999\%) and Sb (Alfa Aesar, 99.9999\%) pieces in an atomic ratio of 1:9 were packed in an Alumina crucible and sealed in a quartz ampoule under partial Ar pressure. The ampoule was heated up to 650 $\celsius$, kept at that temperature for four hours, then slowly cooled down to 620 $\celsius$, after which the excess flux was removed in a centrifuge. The as-grown crystals were thin plates with a typical dimension of 5$\times$5$\times$0.2 $mm^3$. A 1:1:1 HCl-HNO$_3$-H$_2$O solution was used to remove the remnant flux from the surface of the crystals.

Room temperature powder X-ray diffraction (XRD) measurements were carried out on a Rigaku D/Max diffractometer with Cu $K_{\alpha}$ radiation and a graphite monochromator. Rietveld analysis was performed to determine the lattice parameters, using the \textsc{rietica} software package.\cite{rietica}

DC magnetization measurements were performed in a commercial Quantum Design Magnetic Properties Measurement System (QD MPMS) with a He3 insert for temperatures between 0.5 and 2 K. For the plate-like crystals, the shape was assumed to be ellipsoidal, and the demagnetization factor $N_d$ was determined from tabulated values.\cite{Osborn1945} AC magnetization was measured in a dilution refrigerator, using a standard AC susceptometer consisting of two oppositely wound pick-up coils. An external modulation field of 0.1 Oe and 113.7 Hz was applied in the direction parallel to the crystal plate and data were acquired by a lock-in amplifier. After background subtraction, the phase was shifted according to the excitation frequency. In order to obtain absolute values of the magnetization, the data were matched to the results from the QD MPMS measurements. The imaginary part, $\chi ''$, was set to zero at temperatures above $T_c$, using an appropriate offset. The offset and the scaling to absolute values were the same for both the temperature and field sweeps.

Heat capacity of YbSb$_2$ was measured in a QD Physical Property Measurement System (PPMS) with a He3 option, using a thermal relaxation technique. To demonstrate the field dependence of heat capacity, measurements were carried out in a magnetic field applied perpendicular to the crystal plate, with magnitudes ranging from 0 to 90 kOe. Temperature dependent AC resistivity was also measured in the QD PPMS, utilizing a standard four-probe method. The sample was cut into a bar-like shape and four platinum wires were attached to the flat surface using Epo-Tek H20E silver epoxy. An AC current of $\emph{i}$ = 0.1 mA and $f$ = 1000 Hz was applied along the in-plane direction and resistivity data were taken during cooling.

In-plane magnetic penetration depth $\Delta\lambda(T)$ was determined  using a self-resonating tunnel-diode oscillator (TDR)\cite{Degrift1975}, operating at 16 MHz with an amplitude of $H_{AC}\approx 10$ mOe, temperatures down to 50 mK and in static magnetic fields up to $H_{DC}=400$ Oe. The sample was mounted on a sapphire rod with the crystal plate perpendicular to both $H_{AC}$ and $H_{DC}$. Placing the sample into the inductor causes the shift of the resonant frequency $\Delta f(T)=-~G4\pi\chi(T)$, where $G$ is a calibration constant determined by physically pulling the sample out of the coil. With the characteristic sample size $R$, $\Delta \lambda$ can be obtained from $4\pi\chi=(\lambda/R)\tanh (R/\lambda)-1$.\cite{Prozorov2000,Prozorov2006}

\section{Results and Discussion}
\begin{figure}
  \includegraphics[width=\columnwidth]{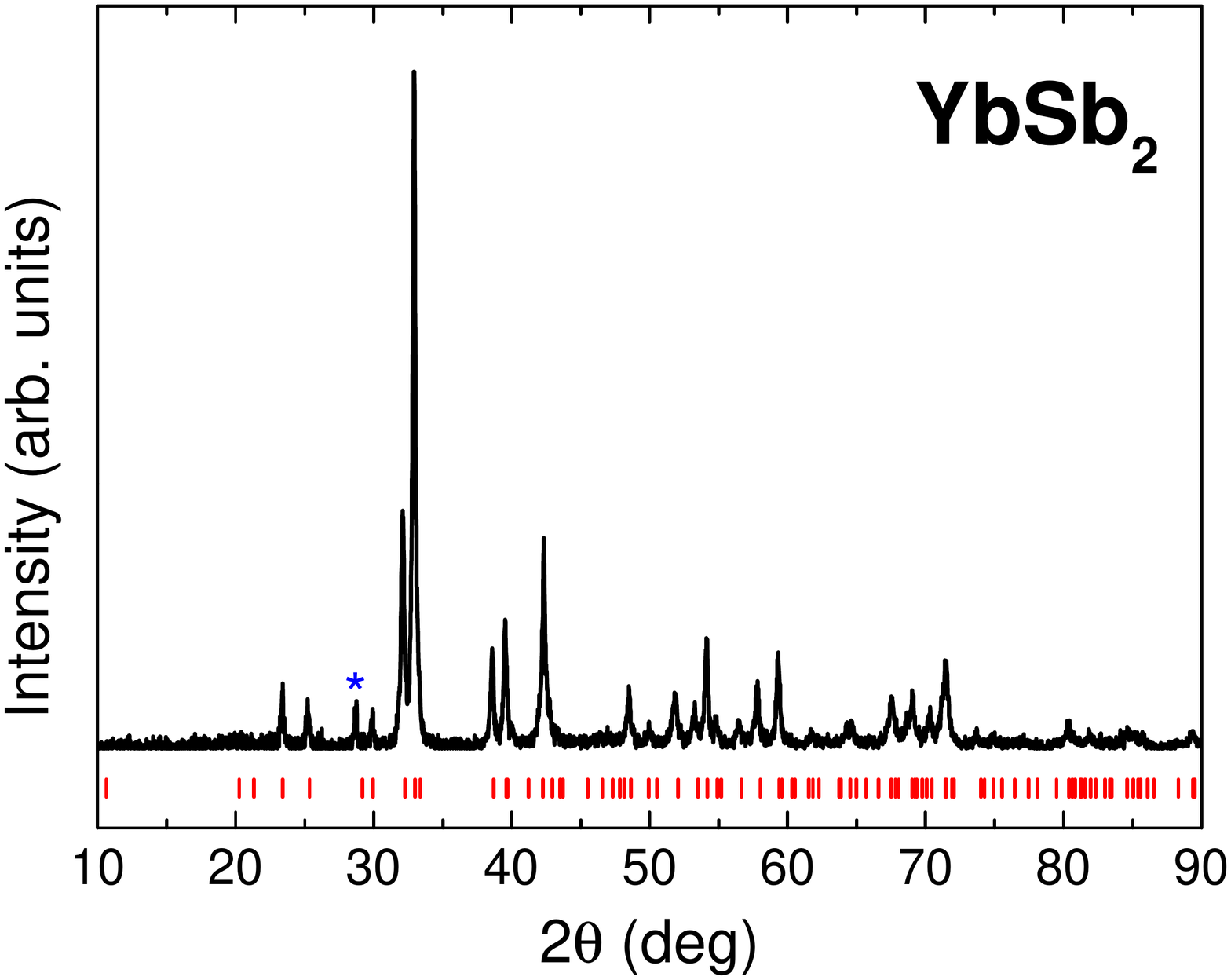}\\
  \caption{Powder XRD pattern of YbSb$_2$ (black line), with peak positions (vertical marks) calculated based on space group $Cmcm$ and lattice parameters $a$ = 4.554 $\AA$, $b$ = 16.715 $\AA$ and $c$ = 4.267 $\AA$. A small amount of residual Sb flux is marked by a blue asterisk.}\label{XRD}
\end{figure}

The powder X-ray diffraction pattern of YbSb$_2$ is shown in Fig. \ref{XRD}. The pattern was refined using a ZrSi$_2$-type orthorhombic structure with space group $Cmcm$ and lattice parameters $a$ = 4.554 $\AA$, $b$ = 16.715 $\AA$ and $c$ = 4.267 $\AA$, in good agreement with the previously reported values.\cite{Wang} A small amount of remnant Sb flux ($\sim$ 5\%) was found and is marked by a blue asterisk in Fig. \ref{XRD}.

\begin{figure}
  \includegraphics[width=\columnwidth]{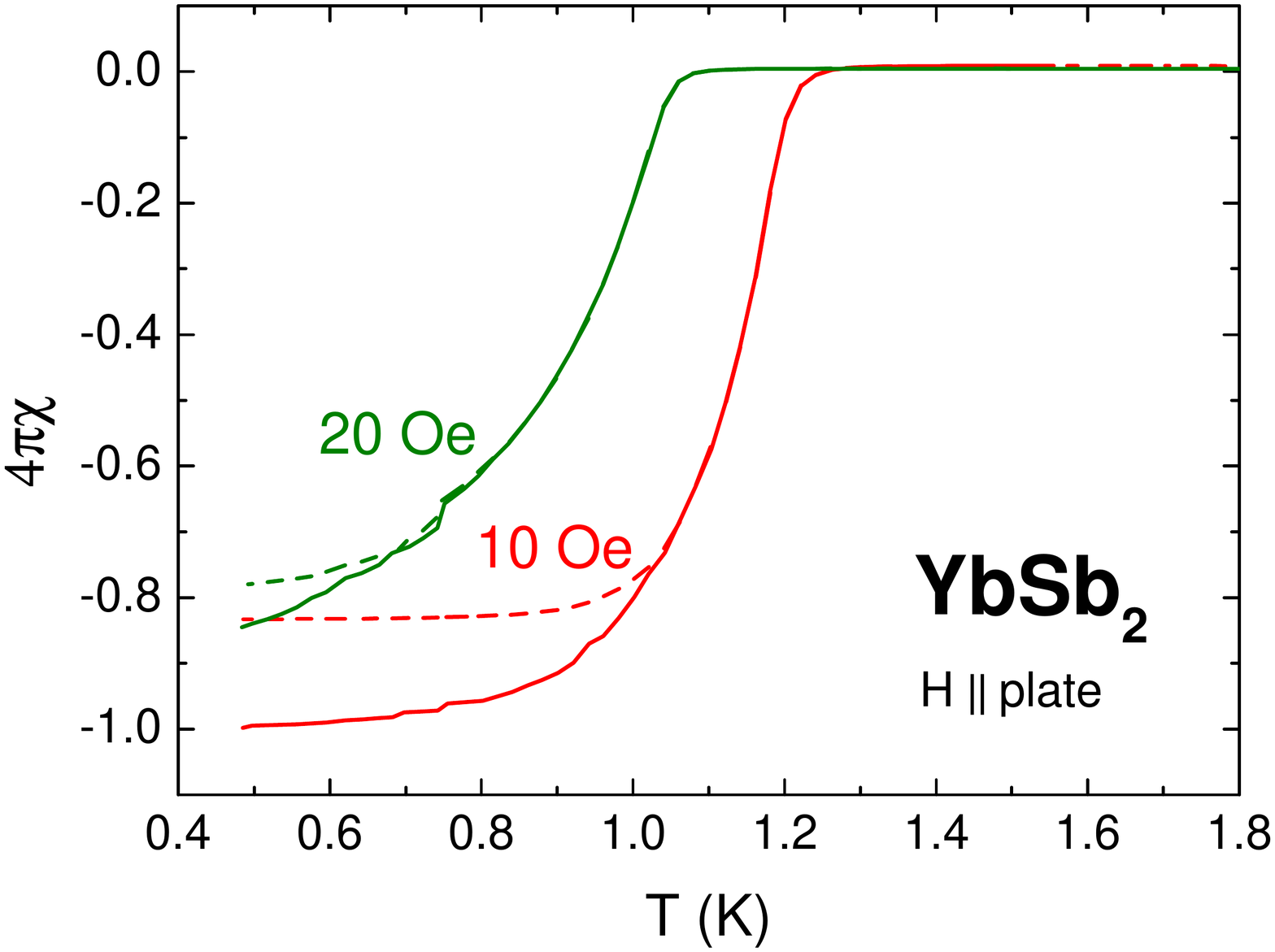}\\
  \caption{Temperature dependent susceptibility of YbSb$_2$, measured in a DC field parallel to the crystal plate. The ZFC and FC data are plotted as solid and dashed lines, respectively.}\label{DCMT}
\end{figure}

\begin{figure}
  \includegraphics[width=0.8\columnwidth]{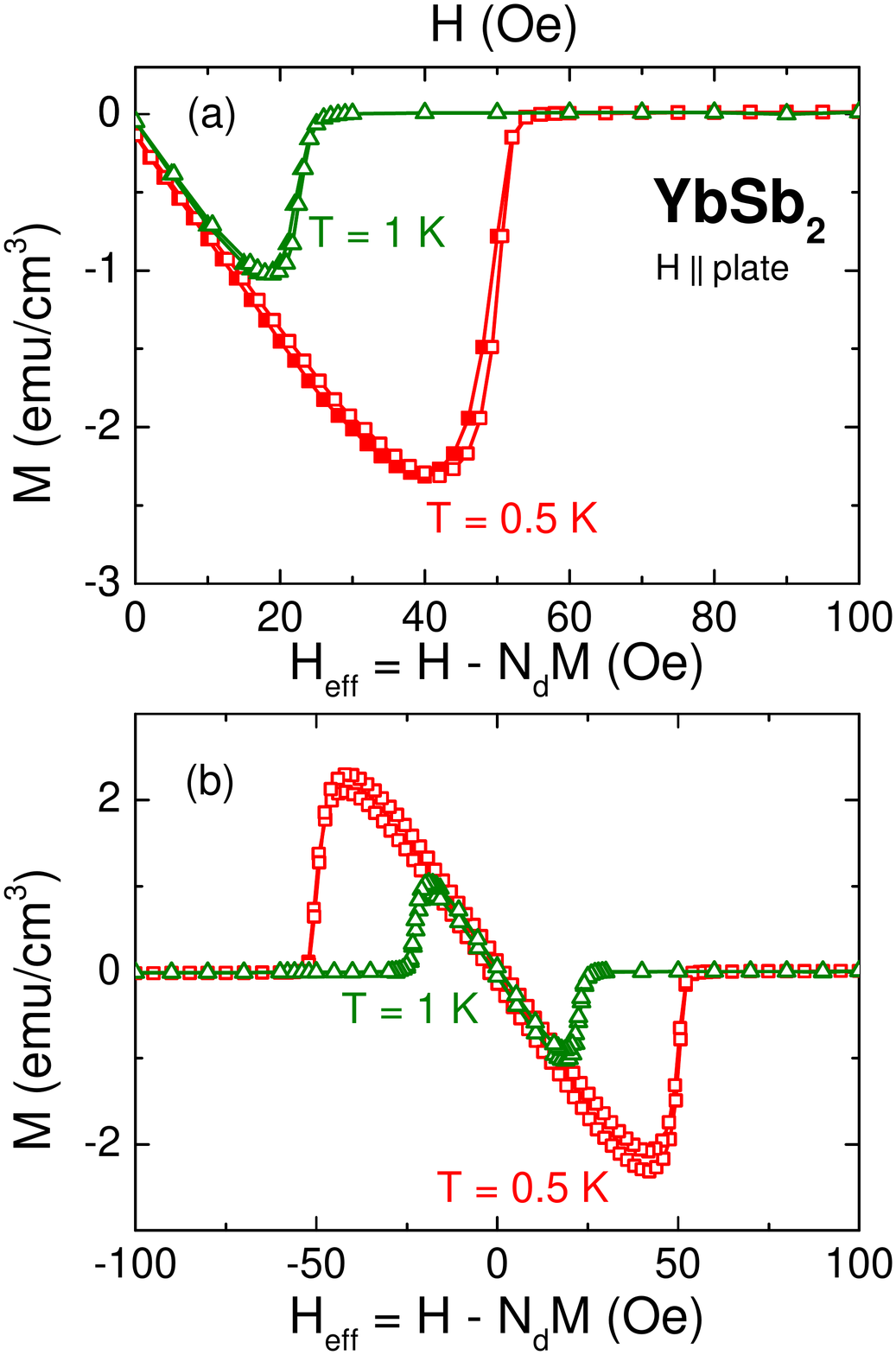}\\
  \caption{(a) Isothermal magnetization and (b) full $M(H)$ loops of YbSb$_2$ for $T$ = 0.5 K (squares) and 1 K (triangles) in $H\parallel$ plate. Solid and open symbols correspond to data before (top axis) and after (bottom axis) the demagnetization correction, respectively (see text).}\label{DCMH}
\end{figure}

Fig. \ref{DCMT} shows the temperature dependent DC magnetic susceptibility $\chi$ of YbSb$_2$, measured in a field $H$ parallel to the crystal plate for $H=$ 10 and 20 Oe. The demagnetizing effect has been taken into account by calculating $\chi=\chi_{measured}/(1-N_{d}\chi_{measured})$, where $\chi_{measured}=M/H$ is the ratio of the measured magnetization $M$ and the applied magnetic field $H$. The demagnetization factor $N_d$ is estimated to be 0.07, if we approximate the shape of the plate-like crystal as an ellipsoid.\cite{Osborn1945} For both zero field cooled (ZFC, solid lines) and field cooled (FC, dashed lines) data, the low temperature susceptibility shows a clear Meissner signal at temperatures below $T_c$ = 1.25 K for $H$ = 10 Oe. The superconducting volume fraction estimated from the ZFC data at this field value is very close to 100\%, indicative of bulk superconductivity. As expected, $T_c$ is suppressed by increasing magnetic field. The DC magnetization isotherms $M(H)$ for $T$ = 0.5 and 1.0 K are shown in Fig. \ref{DCMH}a, before (solid symbols) and after (open symbols) the demagnetization correction. It is clear that the corrected $M(H)$ curves show a step-like jump to zero near the critical field, characteristic of type I superconductivity. The full magnetization loops (Fig. \ref{DCMH}b) also have the shape typical of type I superconductors.\cite{HgMH,TaMH,SnMH}

\begin{figure}
  \includegraphics[width=\columnwidth]{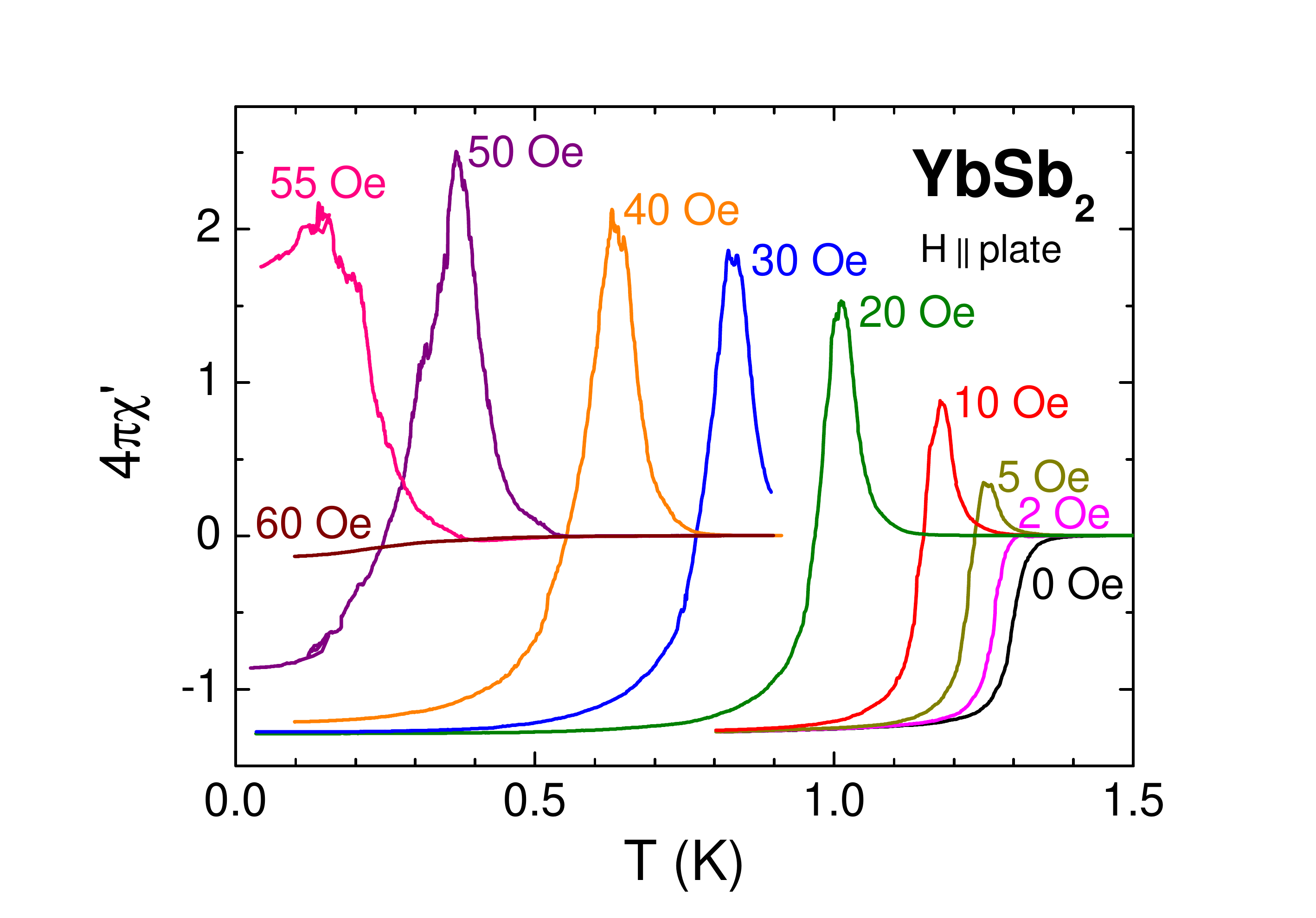}\\
  \caption{Temperature dependent AC susceptibility of YbSb$_2$, measured in fields $H$ up to 60 Oe.}\label{ACMT}
\end{figure}

\begin{figure}
  \includegraphics[width=\columnwidth]{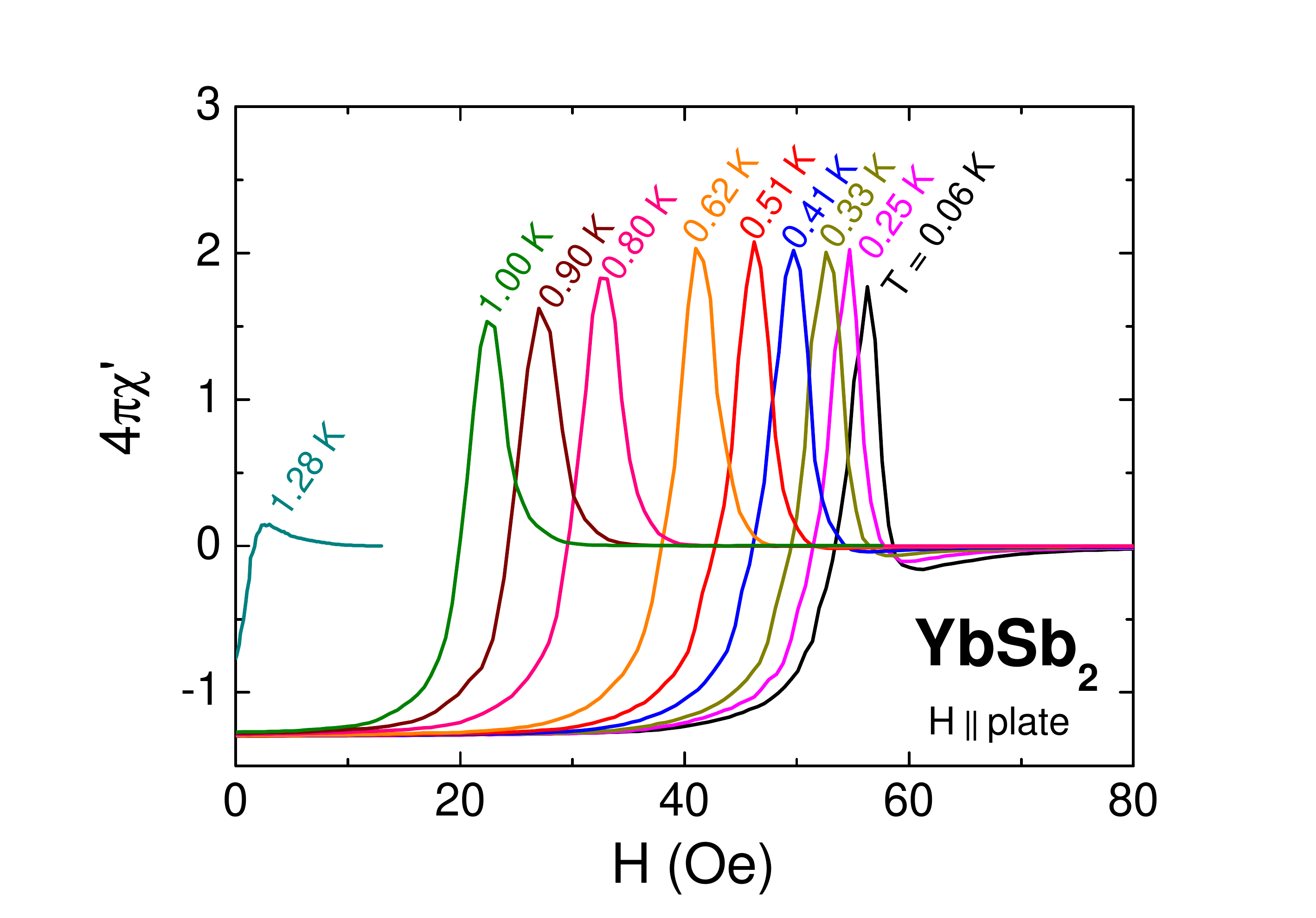}\\
  \caption{Field dependent AC susceptibility of YbSb$_2$, measured for various temperatures from 0.06 K to 1.28 K.}\label{ACMH}
\end{figure}

\begin{figure}
  \includegraphics[width=\columnwidth]{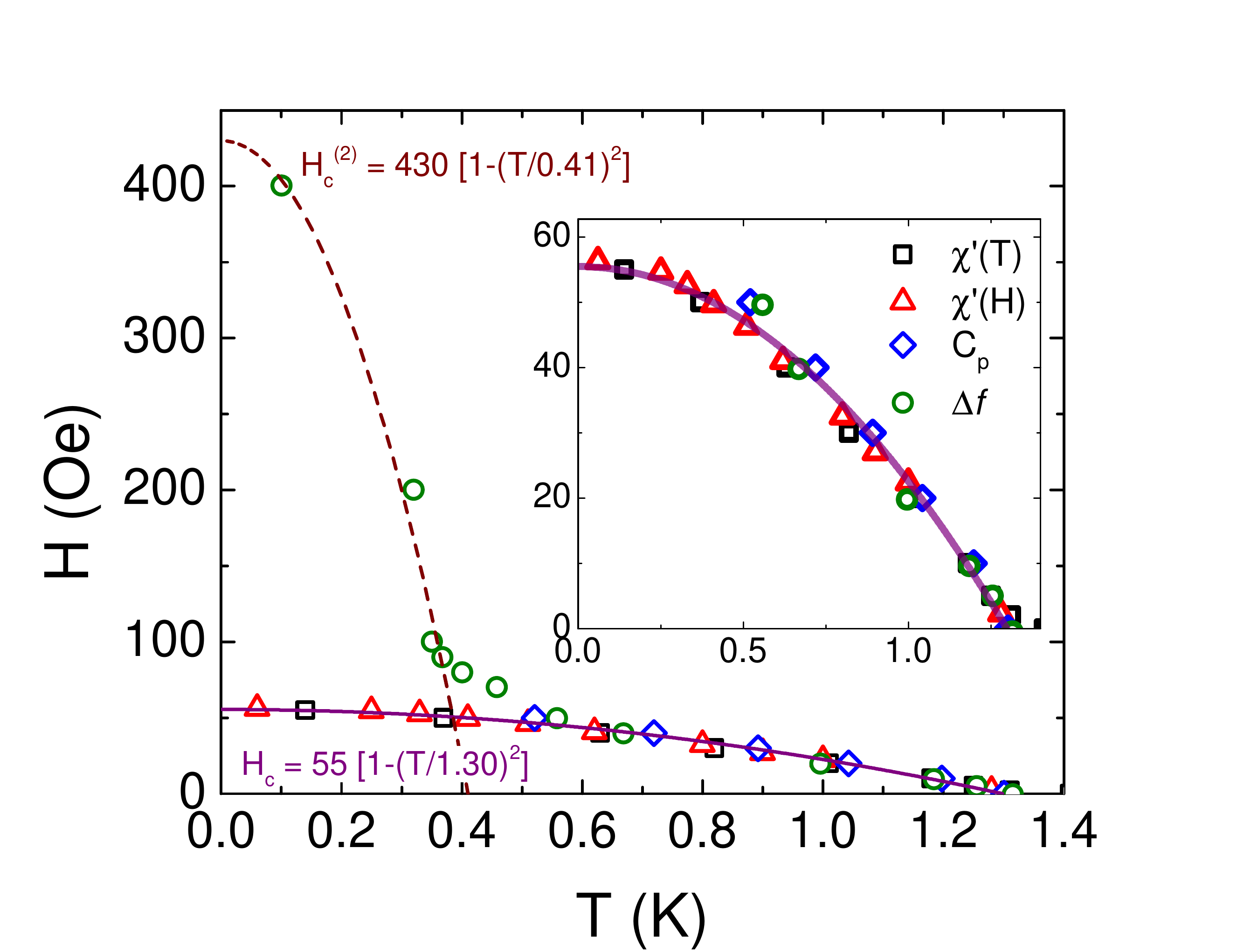}\\
  \caption{$H$-$T$ phase diagram of YbSb$_2$. Points determined from $\chi'(T)$, $\chi'(H)$, $C_{p}(T)$ and $\Delta f(T)$ data are marked by black squares, red triangles, blue diamonds and green circles. The superconductive phase boundary determined from thermodynamic data is illustrated by a purple line. The possible new superconducting state observed in the RF magnetization measurement is marked with a dashed line; inset: a zoomed-in view of the data points from thermodynamic measurements.}\label{PD}
\end{figure}

The AC susceptibility $\chi'$ of YbSb$_2$ as a function of temperature is shown in Fig. \ref{ACMT}. As the field increases from $H$ = 0 to 55 Oe, the onset temperature of the Meissner signal drops from 1.41 K to 0.14 K. In a $H$ = 60 Oe field, the superconductivity is further suppressed and cannot be detected down to 0.06 K. Similarly, a suppression of $H_c$ with $T$ from 56.3 Oe ($T$ = 0.06 K) to 2.3 Oe ($T$ = 1.28 K) is illustrated by the $\chi'(H)$ data in Fig. \ref{ACMH}. The onset critical temperature $T_c$ and critical field $H_c$ values from AC susceptibility measurements are summarized in a $H-T$ phase diagram in Fig. \ref{PD}, and will be discussed later. A noteworthy feature of the $\chi'(T)$ and $\chi'(H)$ curves is the positive peak in the vicinity of the superconducting transition for $H>0$, known as the differential paramagnetic effect (DPE) in superconductors.\cite{DPE} The DPE signal originates from the positive $\partial M/\partial H$ values at temperatures right below $H_c$, and can be observed in either type I or type II superconductors. Nevertheless, since the height of the DPE peak in type II superconductors cannot exceed the absolute value of the diamagnetic susceptibility,\cite{Ag5Pb2O6} the observed large DPE signal (Figs. \ref{ACMT} and \ref{ACMH}) clearly points to type I superconductivity in YbSb$_2$.

\begin{figure}
  \includegraphics[width=\columnwidth]{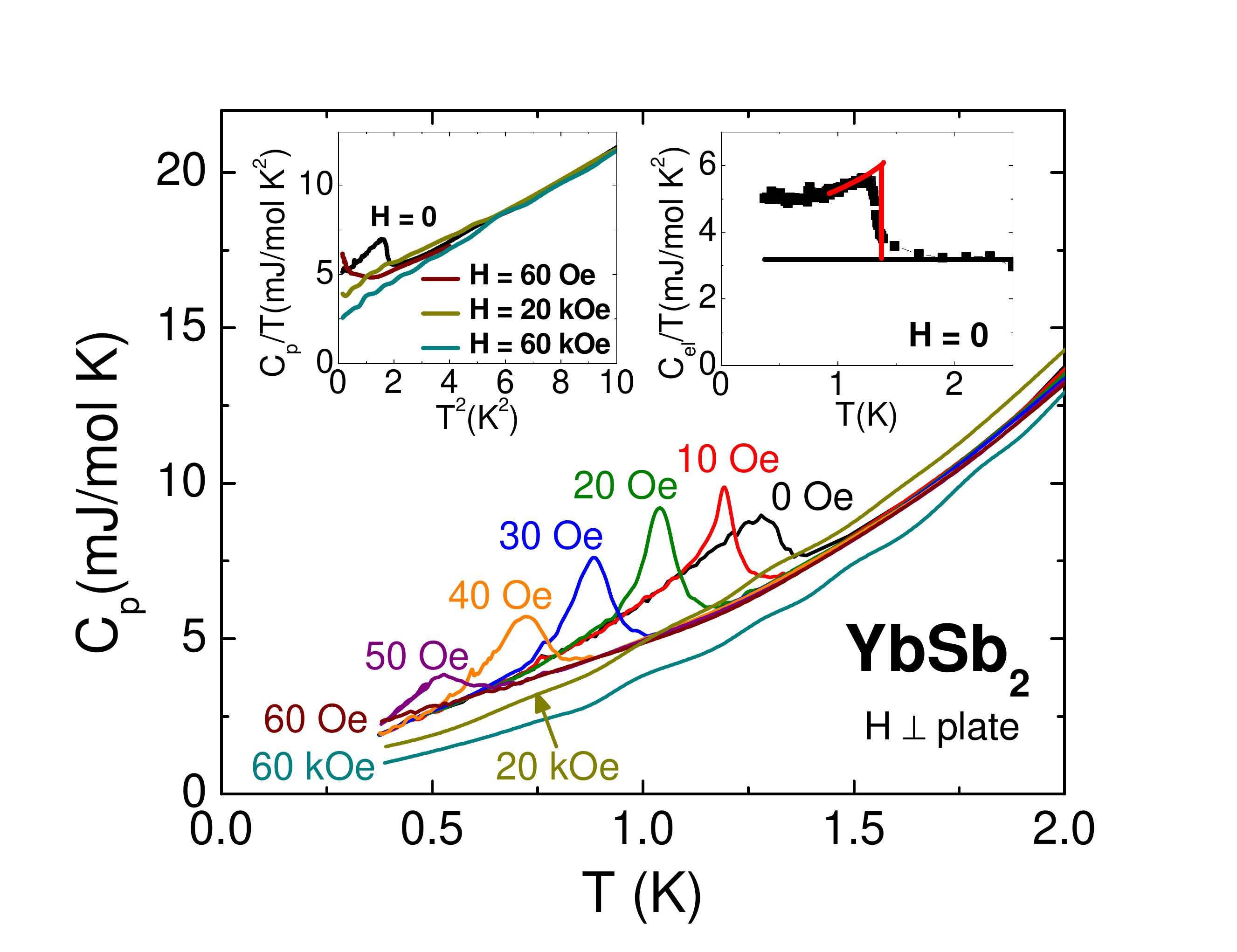}\\
  \caption{Temperature dependent heat capacity of YbSb$_2$ for $H=$ 0, 10, 20, 30, 40, 50, 60 Oe and 20, 60 kOe; left inset: $C_{p}/T$ vs. $T^2$ for $H=0$, 60 Oe, 20 kOe and 60 kOe, showing a field dependent Sommerfeld coefficient $\gamma$; right inset: the superconducting (squares) and normal (black line) electronic specific heat for $H=0$, plotted as $C_{el}/T$ vs. $T$. An entropy conservation construct (red lines) is used to determine the jump in the electronic specific heat at $T_c$.}\label{Cp}
\end{figure}

The field-dependence of the superconducting transition of YbSb$_2$ is further confirmed by heat capacity measurements in fields up to $H$ = 90 kOe. For clarity, a subset of these data are shown in Fig. \ref{Cp}. In the $H=0$ curve, a clear jump in the heat capacity confirms the bulk superconductivity. The transition temperature $T_{c}$, determined at the midpoint of this jump, is close to 1.32 K and agrees well with the previous report.\cite{YamaguchiYbSb2} As $H$ increases, $T_{c}$ monotonically decreases and drops below 0.4 K (the minimum available temperature for these measurements) at $H$ = 60 Oe. The peak at the transition also becomes sharper and higher for $H=$ 10 Oe, compared to that for $H=0$, indicating a change from second to first order phase transition, commonly seen in type I superconductors. As the field is further increased, a non-monotonic change of the electronic and phonon specific heat coefficients $\gamma$ and $\beta$ is revealed by the $C_p/T$ vs. $T^2$ plots (Fig. \ref{Cp}, left inset). While this field dependence remains to be understood, it makes it difficult to determine the electronic specific heat jump $\Delta C_{el}=C_{el,s}-C_{el,n}$ at $T_c$. An additional difficulty in estimating $\Delta C_{el}$ is a possible second superconducting transition at lower temperatures, which will be discussed below. We therefore use the $H=0$ $C_p/T$ vs. $T^2$ data (black, Fig. \ref{Cp}, left inset) to estimate $\gamma$ and $\beta$ from the linear fit. This gives $\gamma\approx$ 3.18 mJ/mol K$^{2}$ and $\beta\approx$ 0.90 mJ/mol K$^{4}$. After subtracting the phonon contribution $C_{ph}=\beta T^{3}$, the jump in the specific heat at the superconducting transition is estimated to be $\Delta C_{el}/\gamma T_{c}~\approx$ 0.85 (Fig. \ref{Cp}, right inset), smaller than the BCS value of 1.43. As already mentioned, this could be due to a second superconducting transition and the non-monotonic field dependence of $\gamma$ and $\beta$, as illustrated by a subset of $C_{p}/T(T^{2})$ curves shown in left inset, Fig. \ref{Cp}. The Debye temperature can also be estimated, using $\theta_{D} = (12\pi ^{4}N_{A}rk_{B}/5\beta )^{1/3}$, where $r=3$ is the number of atoms per formula unit. This yields $\theta_{D}~\approx$ 186 K.

\begin{figure}
  \includegraphics[width=\columnwidth]{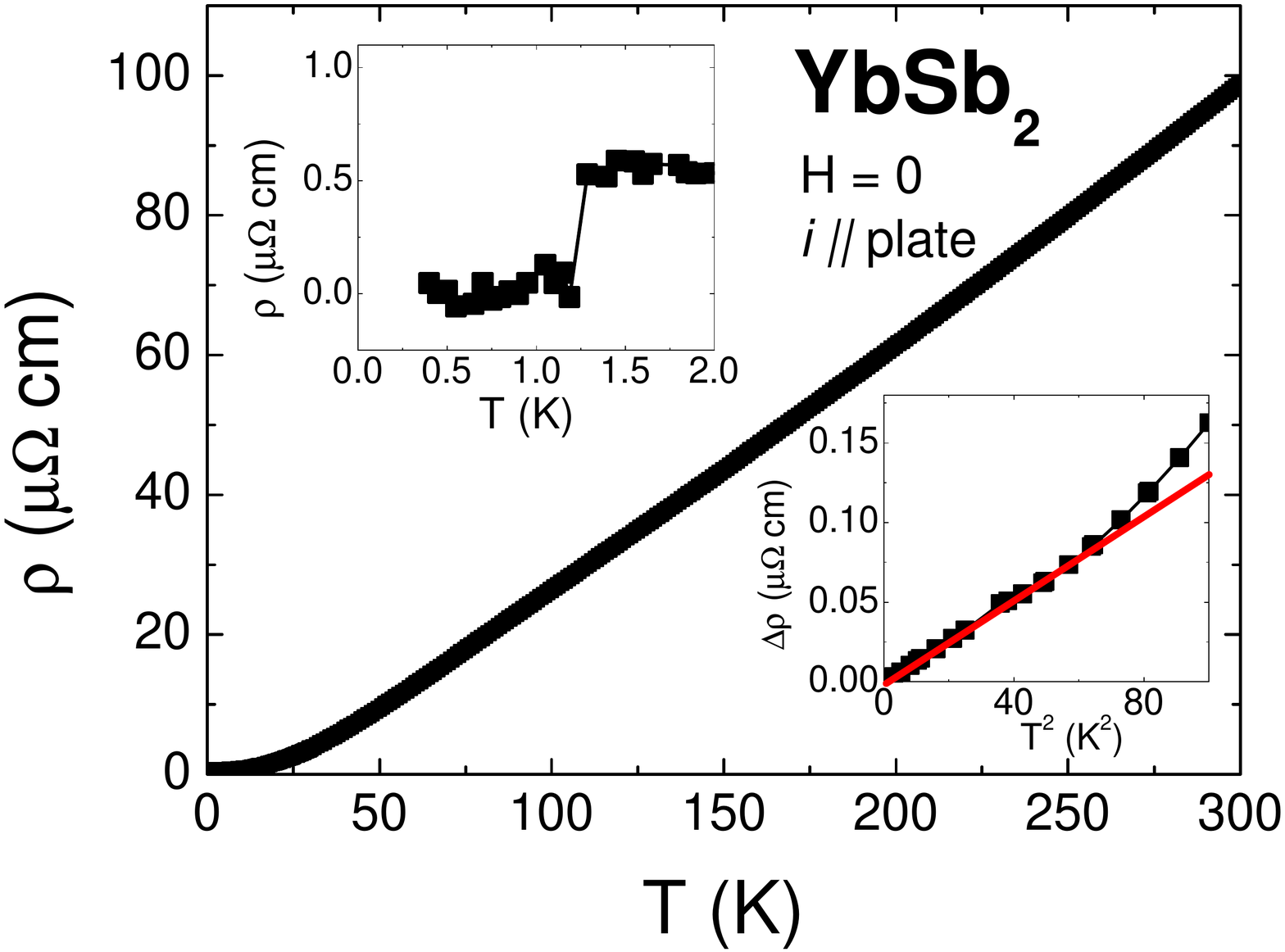}\\
  \caption{AC resistivity of YbSb$_2$, for $H$ = 0 and $i\parallel$ plate. Top inset: a zoomed-in view of the low temperature resistivity around $T_c$; bottom inset: $\Delta\rho=\rho-\rho_{0}$ vs. $T^2$ (symbols), together with a linear fit (solid line).}\label{R}
\end{figure}

The AC resistivity of YbSb$_2$, for $H=0$ and $i\parallel$ plate, is shown in Fig. \ref{R}. At high temperatures, a linear temperature dependence of $\rho(T)$ is evident, as expected for metals. As seen in the upper inset, the resistivity drops to zero at $T_c$ = 1.25 K, with a residual resistivity $\rho_{0}(2K)=0.53~\mu\Omega cm$ just above $T_c$. The residual resistivity ratio, calculated as $RRR=\rho(300K)/\rho(2K)$, is around 186, indicative of a good quality metal. At temperatures below 8 K and above $T_c$, the resistivity shows a quadratic dependence on temperature: $\rho=\rho_{0}+AT^{2}$. From a linear fit of $\Delta\rho=\rho-\rho_{0}$ vs. $T^{2}$, the coefficient $A$ is determined to be $0.0013~\mu\Omega$ cmK$^{-2}$. The Kadowaki-Woods (KW) ratio $A/\gamma^{2}=12.8a_{0}$, where $a_{0}=10^{-5}\mu\Omega$ cm mol$^{2}$K$^{2}$mJ$^{-2}$ is a nearly universal value observed in strongly correlated electron systems. \cite{KW2} This large KW ratio is consistent with the analogous value previously reported \cite{SatodHvA}, and could be associated with electron-phonon scattering or enhanced electron correlations.

Based on the resistivity and heat capacity data, several superconducting parameters, such as the London penetration depth $\lambda_L$, coherence length $\xi$, Ginzburg-Landau parameter $\kappa$ and electron-phonon coupling constant $\lambda_{el-ph}$ can be estimated. With four formula units per unit cell volume ($V$), the electron density $n$ of YbSb$_2$ can be calculated as $n=8/V=2.483\times10^{-2}\AA^{-3}$, assuming the valence of Yb to be 2+. The Fermi wave vector $k_F$ can be roughly estimated if we assume a spherical Fermi surface, which gives $k_{F}=(3n\pi^{2})^{\frac{1}{3}}=0.903~\AA^{-1}$. The Fermi wave vector $k_{F}$, together with the Sommerfeld coefficient $\gamma=$ 3.18 mJ/molK$^{2}$ = 6.56$\times$10$^{-5}$ J/cm$^{3}$K$^{2}$, yield the effective electron mass $m^{*}=\hbar^{2}k_{F}^{2}\gamma/\pi^{2}nk_{B}^{2}=1.39~m_{0}$, where $m_0$ is the free electron mass. The London penetration depth $\lambda_{L}(0)$ and coherence length $\xi(0)$ can also be derived as $\lambda_{L}(0)=(m^{*}/\mu_{0}ne^{2})^{1/2}=$ 40 nm and $\xi(0)=0.18\hbar^{2}k_{F}/k_{B}T_{c}m^{*}=$ 826 nm. It results that the GL parameter $\kappa=\lambda_{L}(0)/\xi(0)=0.05\ll1/\sqrt{2}$, confirming the type I superconductivity in YbSb$_2$. According to the McMillan theory,\cite{McMillan} the electron-phonon coupling is given by \begin{center}
    $\lambda_{el-ph}=\frac{1.04+\mu^{*}\ln(\theta_{D}/1.45T_{c})}{(1-0.62\mu^{*})\ln(\theta_{D}/1.45T_{c})-1.04}$
\end{center}
where the Coulomb pseudopotential $\mu^{*}$ is usually between 0.1 and 0.15. Assuming $\mu^{*}=0.13$, the electron-phonon coupling is estimated to be $\lambda_{el-ph}\approx0.51$, suggesting that YbSb$_2$ is a weakly coupled BCS superconductor. Moreover, the $\lambda_{el-ph}$ value confirms the effective electron mass $m^{*}$ as calculated using $m^{*}=(1+\lambda_{el-ph})m_{0}$ which gives $m^{*}=1.51m_{0}$.

\begin{figure}
  \includegraphics[width=\columnwidth]{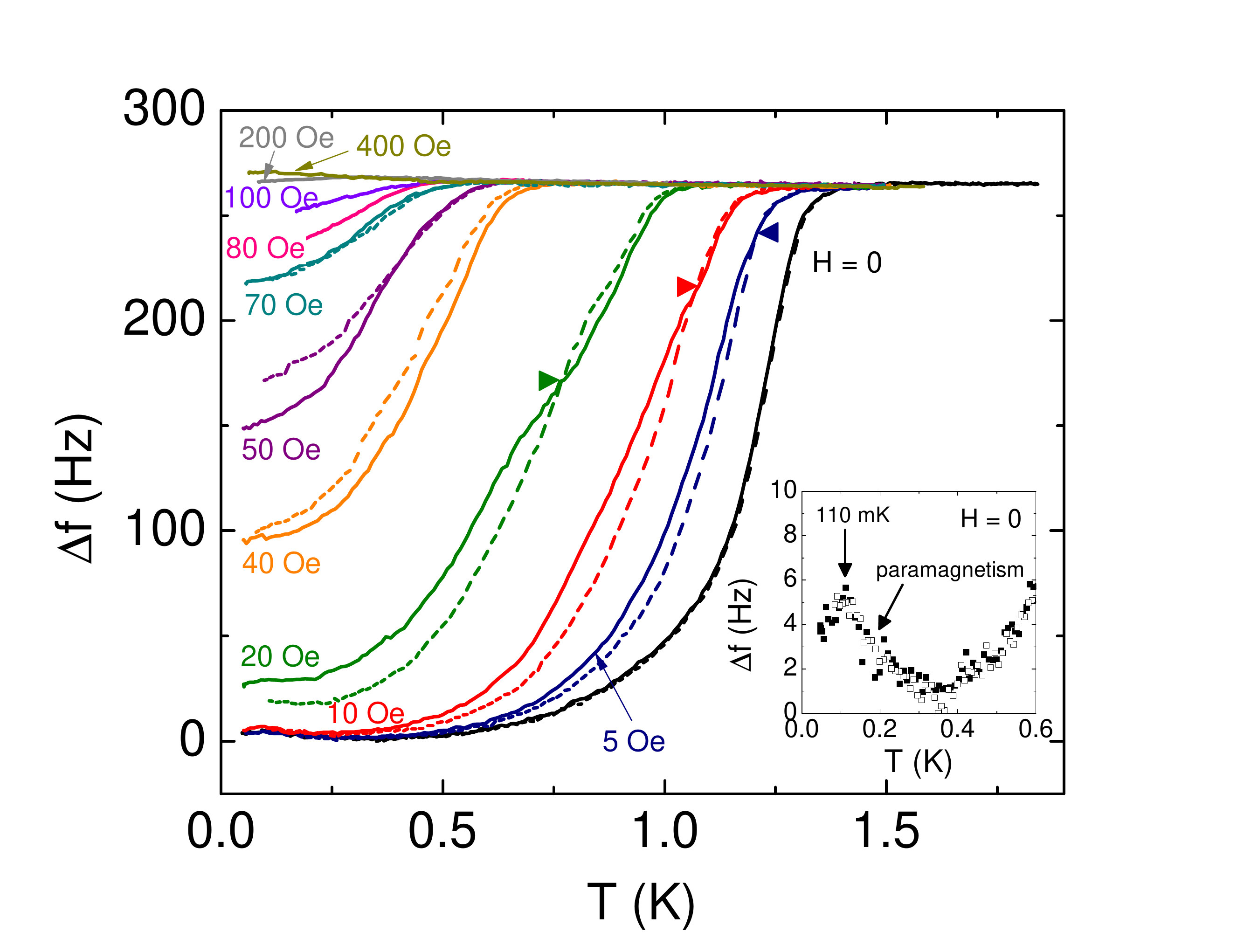}\\
  \caption{Shift of resonant frequency $\Delta f=-G4\pi\chi(T)$, measured in fields up to 400 Oe. The ZFC and FC data are shown as solid and dashed lines, respectively; inset: a zoomed-in view of the low temperature data for $H=$ 0, with ZFC and FC data shown as solid and open symbols.}\label{AmesFig1}
\end{figure}

\begin{figure}
  \includegraphics[width=\columnwidth]{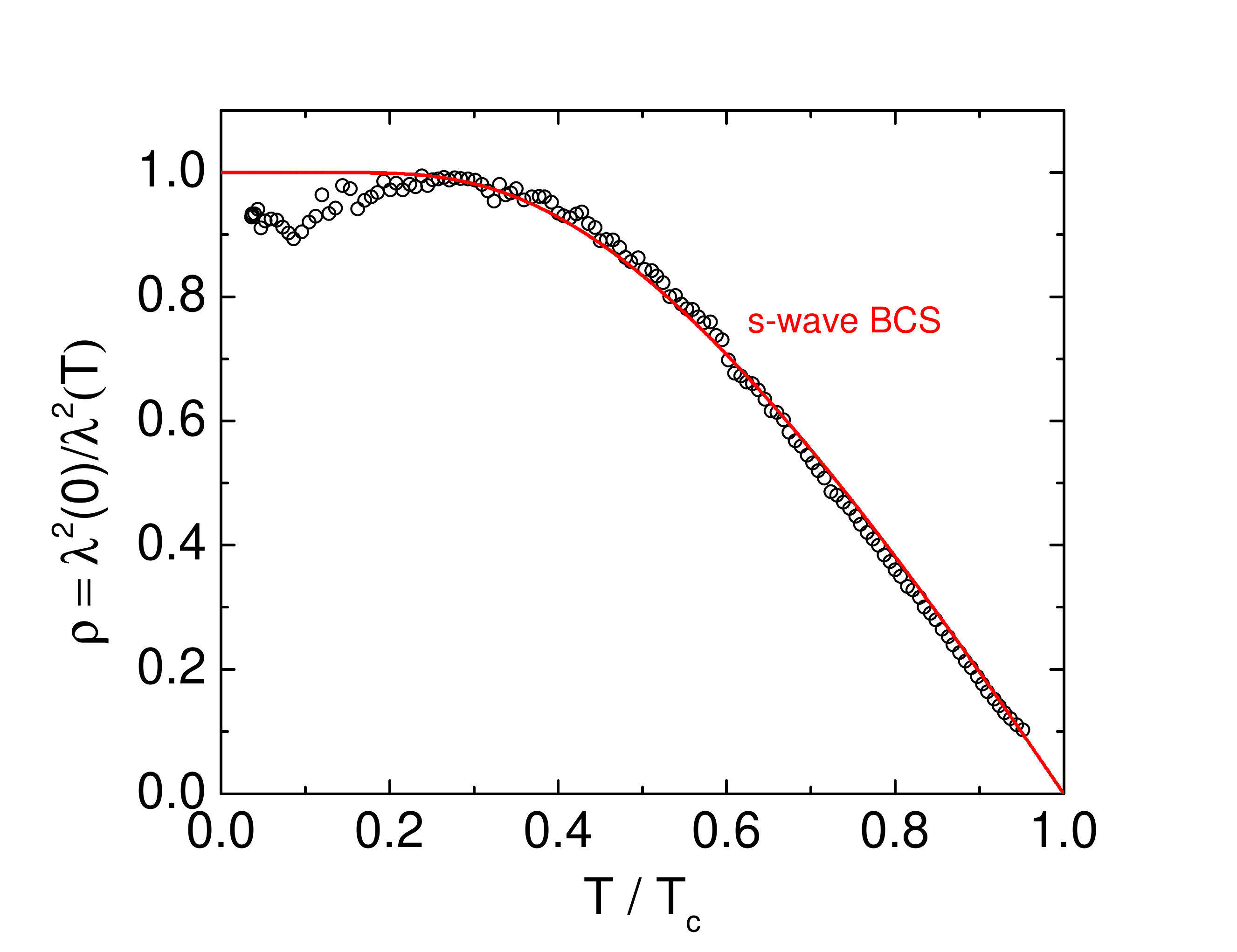}\\
  \caption{Superfluid density calculated as $\rho=\lambda^{2}(0)/\lambda^{2}(T)$. The solid red line shows a fit to the typical behavior of an $s$-wave BCS superconductor.}\label{AmesFig2}
\end{figure}

Figure \ref{AmesFig1} shows frequency $\Delta f (T) \sim \chi (T)$ measured for a $0.7\times0.5\times0.3$ mm$^3$ sample. Data above $T_c$ represent a combination of magnetic and resistive responses in the normal state of YbSb$_2$. The skin depth at $T=1.5$ K was estimated to be $\delta\approx 8.9~\mu m$, which is much smaller than any dimension of the sample. The skin depth $\delta$ was calculated with the residual resistivity $\rho_0=$ 0.53 $\mu\Omega$cm and operating frequency of 16 MHz. A slight upturn before the superconducting transition can be attributed to the response of some paramagnetic impurities. $T_c$ was determined as the temperature where $\Delta f (T)$ deviates from the normal state behavior.

In the pure Meissner state (Fig. \ref{AmesFig1}), for $H_{DC}=0$, both ZFC and FC curves coincide. Additionally, apart from the superconducting transition, a small feature near 0.11 K was observed, as shown in the inset. This may be attributed to a phase associated with extrinsic magnetic impurities. In Fig. \ref{AmesFig2}, the calculated superfluid density, $\rho=\lambda^2(0)/\lambda^2(T)$, is found to be consistent with a single-gap $s$-wave pairing in YbSb$_2$, except for the impurity contribution which modifies the curve at the lowest temperatures. In the presence of magnetic impurities with magnetic permeability $\mu(T)$, the actual measured penetration depth is renormalized as $\lambda_{m}=\sqrt{\mu(T)}\lambda$, so that the experimentally constructed superfluid density is given by $\rho_{m}=\lambda^{2}(0)/[\mu(T)\lambda^{2}(T)]$. For paramagnetic impurities behaving following the Curie-law, this renormalization leads to a downward trend of $\rho_{m}$ in the region where real superfluid density is already flat. At higher temperatures the contribution of paramagnetic impurities rapidly vanishes, since $\mu=1+4\pi\chi=1+C/T$, where $C$ is the Curie constant.

\begin{table}
  \centering
  \caption{Measured and calculated superconducting and thermodynamic parameters of YbSb$_2$.}  \label{para}
  \renewcommand{\arraystretch}{1.2}
  \begin{tabular}{p{0.5\columnwidth} p{0.25\columnwidth} p{0.25\columnwidth}}
    \hline\hline
    $T_c$ (K) & & 1.30 $\pm$ 0.2\\
    $H_{c}(0)$ (Oe) & & 55 $\pm$ 2 \\
    $\gamma$ (mJ/mol K$^{2}$) & & 3.18 \\
    $\beta$ (mJ/mol K$^{4}$) & & 0.90\\
    $\theta_{D}$ (K) & & 186\\
    $\Delta C_{el}/\gamma T_{c}$ & & 0.85 \\
    $RRR$ & & 186 \\
    $A$ ($\mu\Omega$cm/K$^{2}$) & & 0.0013 \\
    $A/\gamma^{2}$ (10$^{-5}\mu\Omega$mol$^2$K$^2$mJ$^{-2}$) & & 12.8 \\
    $k_F$ ($\AA^{-1}$) & & 0.903 \\
    $m^{*}$ ($m_0$) & & 1.45 $\pm$ 0.06 \\
    $\lambda_{L}(0)$ (nm) & & 40 \\
    $\xi(0)$ (nm) & & 826 \\
    $\kappa$ & & 0.05 \\
    $\lambda_{el-ph}$ & & 0.51 \\
    $T_c^{(2)}$ (K) & & 0.41 \\
    $H_{c}^{(2)}(0)$ (Oe) & & 430 \\
  \hline\hline
  \end{tabular}
\end{table}

In non-zero magnetic field, ZFC-FC $\Delta f$ curves (Fig. \ref{AmesFig1}) show hysteresis up to $H=$ 50 Oe. Interestingly, the FC data indicate stronger repulsion below an intermediate temperature, marked with solid triangles, which systematically decreases with increasing $H$. This crossover no longer exists above 40 Oe, and ZFC data recover stronger repulsion. Above 70 Oe, which is greater than $H_c$ determined by various thermodynamic measurements, data still show diamagnetic response without hysteresis, which persists up to 400 Oe. It should also be noted that the AC susceptibility for $H$ = 60 Oe (Fig. \ref{ACMT}) remains slightly diamagnetic after the DPE peak vanishes, consistent with the results displayed in Fig. \ref{AmesFig1}. The origin of the diamagnetism above the bulk $H_c$ is not clear, however the $H$-$T$ phase diagram shown in Fig. \ref{PD} is reminiscent of a field-dependent pairing state with multiple order parameters in the heavy fermion superconductor PrOs$_4$Sb$_{12}$. \cite{Cichorek2005,Maple2007} More detailed measurements are required to fully understand the rich physics of this unconventional behavior, and to clarify whether this could be bulk or surface superconductivity.

The relationship between $T_c$ and $H_c$ is summarized in the $H$-$T$ phase diagram (Fig. \ref{PD}). The $H_c$ values determined from $\chi'(T)$ (squares), $\chi'(H)$ (triangles), $C_p$ (diamonds) and $\Delta f$ (circles) below $\sim$ 80 Oe can be fit to the expected BCS temperature dependence $H_{c}(T)=H_{c}(0)[1-(T/T_{c})^{2}]$ (solid line, Fig. \ref{PD} inset). This gives $H_{c}(0)=$ 55 Oe and $T_{c}=1.30~K$. The possible new superconducting state inferred from the RF magnetization (Fig. \ref{AmesFig1}) can also be described with a similar equation $H_{c}^{(2)}(T)=H_{c}^{(2)}(0)[1-(T/T_{c}^{(2)})^2]$ (dashed line, Fig. \ref{PD}), which gives $H_{c}^{(2)}(0)=$ 430 Oe and $T_{c}^{(2)}=$ 0.41 K.

The superconducting and thermodynamic parameters of YbSb$_2$ are summarized in Table \ref{para}. Several traits of type I superconductors have been observed in this compound, including a small GL parameter $\kappa$ = 0.05, typical shape of the  DC $M(H)$ isotherms (Fig. \ref{DCMT}), a strong DPE signal in the AC magnetization (Fig. \ref{ACMH}), small critical field values, and a change from second to first order phase transition induced by magnetic field and visible in specific heat data (Fig. \ref{Cp}). All these observations provide proof of the type I superconductivity in YbSb$_2$. In addition, a possible second superconducting state at lower temperatures is observed in RF magnetization (Fig. \ref{AmesFig1}), which reveals unconventional behavior, as of yet not fully understood. This calls for more experiments to elucidate the underlying physics in YbSb$_2$.

\section{ACKNOWLEDGMENTS}
Work at Rice University was supported by AFOSR MURI. Work at Ames was supported by the U.S. Department of Energy, Office of Basic Energy Sciences, Division of Materials Sciences and Engineering under contract No. DE-AC02-07CH11358. The authors thank Yuri Prots and Juri Grin for help with the structural characterization.


\begin{thebibliography}{24}
\expandafter\ifx\csname natexlab\endcsname\relax\def\natexlab#1{#1}\fi
\expandafter\ifx\csname bibnamefont\endcsname\relax
  \def\bibnamefont#1{#1}\fi
\expandafter\ifx\csname bibfnamefont\endcsname\relax
  \def\bibfnamefont#1{#1}\fi
\expandafter\ifx\csname citenamefont\endcsname\relax
  \def\citenamefont#1{#1}\fi
\expandafter\ifx\csname url\endcsname\relax
  \def\url#1{\texttt{#1}}\fi
\expandafter\ifx\csname urlprefix\endcsname\relax\def\urlprefix{URL }\fi
\providecommand{\bibinfo}[2]{#2}
\providecommand{\eprint}[2][]{\url{#2}}

\bibitem[{\citenamefont{Kobayashi and Tsujikawa}(1981)}]{CxK}
\bibinfo{author}{\bibfnamefont{M.}~\bibnamefont{Kobayashi}} \bibnamefont{and}
  \bibinfo{author}{\bibfnamefont{I.}~\bibnamefont{Tsujikawa}},
  \bibinfo{journal}{J. Phys. Soc. Jpn.} \textbf{\bibinfo{volume}{50}},
  \bibinfo{pages}{3245} (\bibinfo{year}{1981}).

\bibitem[{\citenamefont{Palstra et~al.}(1986)\citenamefont{Palstra, Lu,
  Menovsky, Nieuwenhuys, Kes, and Mydosh}}]{RPd2Si2andRRh2Si2}
\bibinfo{author}{\bibfnamefont{T.~T.~M.} \bibnamefont{Palstra}},
  \bibinfo{author}{\bibfnamefont{G.}~\bibnamefont{Lu}},
  \bibinfo{author}{\bibfnamefont{A.~A.} \bibnamefont{Menovsky}},
  \bibinfo{author}{\bibfnamefont{G.~J.} \bibnamefont{Nieuwenhuys}},
  \bibinfo{author}{\bibfnamefont{P.~H.} \bibnamefont{Kes}}, \bibnamefont{and}
  \bibinfo{author}{\bibfnamefont{J.~A.} \bibnamefont{Mydosh}},
  \bibinfo{journal}{Phys. Rev. B} \textbf{\bibinfo{volume}{34}},
  \bibinfo{pages}{4566} (\bibinfo{year}{1986}).

\bibitem[{\citenamefont{Gottlieb et~al.}(1992)\citenamefont{Gottlieb,
  Lasjaunias, Tholence, Laborde, Thomas, and Madar}}]{TaSi2}
\bibinfo{author}{\bibfnamefont{U.}~\bibnamefont{Gottlieb}},
  \bibinfo{author}{\bibfnamefont{J.~C.} \bibnamefont{Lasjaunias}},
  \bibinfo{author}{\bibfnamefont{J.~L.} \bibnamefont{Tholence}},
  \bibinfo{author}{\bibfnamefont{O.}~\bibnamefont{Laborde}},
  \bibinfo{author}{\bibfnamefont{O.}~\bibnamefont{Thomas}}, \bibnamefont{and}
  \bibinfo{author}{\bibfnamefont{R.}~\bibnamefont{Madar}},
  \bibinfo{journal}{Phys. Rev. B} \textbf{\bibinfo{volume}{45}},
  \bibinfo{pages}{4803} (\bibinfo{year}{1992}).

\bibitem[{\citenamefont{Yonezawa and Maeno}(2005)}]{Ag5Pb2O6}
\bibinfo{author}{\bibfnamefont{S.}~\bibnamefont{Yonezawa}} \bibnamefont{and}
  \bibinfo{author}{\bibfnamefont{Y.}~\bibnamefont{Maeno}},
  \bibinfo{journal}{Phys. Rev. B} \textbf{\bibinfo{volume}{72}},
  \bibinfo{pages}{180504(R)} (\bibinfo{year}{2005}).

\bibitem[{\citenamefont{Anand et~al.}(2011)\citenamefont{Anand, Hillier,
  Adroja, Strydom, Michor, McEwen, and Rainford}}]{LaRhSi3}
\bibinfo{author}{\bibfnamefont{V.~K.} \bibnamefont{Anand}},
  \bibinfo{author}{\bibfnamefont{A.~D.} \bibnamefont{Hillier}},
  \bibinfo{author}{\bibfnamefont{D.~T.} \bibnamefont{Adroja}},
  \bibinfo{author}{\bibfnamefont{A.~M.} \bibnamefont{Strydom}},
  \bibinfo{author}{\bibfnamefont{H.}~\bibnamefont{Michor}},
  \bibinfo{author}{\bibfnamefont{K.~A.} \bibnamefont{McEwen}},
  \bibnamefont{and} \bibinfo{author}{\bibfnamefont{B.~D.}
  \bibnamefont{Rainford}}, \bibinfo{journal}{Phys. Rev. B}
  \textbf{\bibinfo{volume}{83}}, \bibinfo{pages}{064522}
  (\bibinfo{year}{2011}).

\bibitem[{\citenamefont{Svanidze and Morosan}()}]{RGa3}
\bibinfo{author}{\bibfnamefont{E.}~\bibnamefont{Svanidze}} \bibnamefont{and}
  \bibinfo{author}{\bibfnamefont{E.}~\bibnamefont{Morosan}},
  \bibinfo{journal}{in preparation}.

\bibitem[{\citenamefont{Yamaguchi et~al.}(1987)\citenamefont{Yamaguchi, Waki,
  and Mitsuoi}}]{YamaguchiYbSb2}
\bibinfo{author}{\bibfnamefont{Y.}~\bibnamefont{Yamaguchi}},
  \bibinfo{author}{\bibfnamefont{S.}~\bibnamefont{Waki}}, \bibnamefont{and}
  \bibinfo{author}{\bibfnamefont{K.}~\bibnamefont{Mitsuoi}},
  \bibinfo{journal}{J. Phys. Soc. Jpn.} \textbf{\bibinfo{volume}{56}},
  \bibinfo{pages}{419} (\bibinfo{year}{1987}).

\bibitem[{\citenamefont{Sato et~al.}(1999)\citenamefont{Sato, Kinokiri,
  Komatsubara, and Harima}}]{SatodHvA}
\bibinfo{author}{\bibfnamefont{N.}~\bibnamefont{Sato}},
  \bibinfo{author}{\bibfnamefont{T.}~\bibnamefont{Kinokiri}},
  \bibinfo{author}{\bibfnamefont{T.}~\bibnamefont{Komatsubara}},
  \bibnamefont{and} \bibinfo{author}{\bibfnamefont{H.}~\bibnamefont{Harima}},
  \bibinfo{journal}{Phys. Rev. B} \textbf{\bibinfo{volume}{59}},
  \bibinfo{pages}{4714} (\bibinfo{year}{1999}).

\bibitem[{\citenamefont{Shirakawa et~al.}(2000)\citenamefont{Shirakawa, Koiwai,
  and Uwatoko}}]{pressureYbSb2}
\bibinfo{author}{\bibfnamefont{N.}~\bibnamefont{Shirakawa}},
  \bibinfo{author}{\bibfnamefont{S.}~\bibnamefont{Koiwai}}, \bibnamefont{and}
  \bibinfo{author}{\bibfnamefont{Y.}~\bibnamefont{Uwatoko}},
  \bibinfo{journal}{High Pressure Conference of Japan}
  \textbf{\bibinfo{volume}{10}}, \bibinfo{pages}{2} (\bibinfo{year}{2000}).

\bibitem[{\citenamefont{Kohori et~al.}(2003)\citenamefont{Kohori, Kohara, Sato,
  and Kinokiri}}]{SbNQR}
\bibinfo{author}{\bibfnamefont{Y.}~\bibnamefont{Kohori}},
  \bibinfo{author}{\bibfnamefont{T.}~\bibnamefont{Kohara}},
  \bibinfo{author}{\bibfnamefont{N.}~\bibnamefont{Sato}}, \bibnamefont{and}
  \bibinfo{author}{\bibfnamefont{T.}~\bibnamefont{Kinokiri}},
  \bibinfo{journal}{Physica C} \textbf{\bibinfo{volume}{388-389}},
  \bibinfo{pages}{579} (\bibinfo{year}{2003}).

\bibitem[{\citenamefont{Hunter}(1998)}]{rietica}
\bibinfo{author}{\bibfnamefont{B.}~\bibnamefont{Hunter}},
  \bibinfo{journal}{International Union of Crystallography Commission on Powder
  Diffraction Newsletter} p. \bibinfo{pages}{No. 20} (\bibinfo{year}{1998}).

\bibitem[{\citenamefont{Osborn}(1945)}]{Osborn1945}
\bibinfo{author}{\bibfnamefont{J.~A.} \bibnamefont{Osborn}},
  \bibinfo{journal}{Phys. Rev.} \textbf{\bibinfo{volume}{67}},
  \bibinfo{pages}{351} (\bibinfo{year}{1945}).

\bibitem[{\citenamefont{Degrift}(1975)}]{Degrift1975}
\bibinfo{author}{\bibfnamefont{C.~T.~V.} \bibnamefont{Degrift}},
  \bibinfo{journal}{Rev. Sci. Instrum.} \textbf{\bibinfo{volume}{46}},
  \bibinfo{pages}{599} (\bibinfo{year}{1975}).

\bibitem[{\citenamefont{Prozorov et~al.}(2000)\citenamefont{Prozorov,
  Giannetta, Carrington, and Araujo-Moreira}}]{Prozorov2000}
\bibinfo{author}{\bibfnamefont{R.}~\bibnamefont{Prozorov}},
  \bibinfo{author}{\bibfnamefont{R.~W.} \bibnamefont{Giannetta}},
  \bibinfo{author}{\bibfnamefont{A.}~\bibnamefont{Carrington}},
  \bibnamefont{and} \bibinfo{author}{\bibfnamefont{F.~M.}
  \bibnamefont{Araujo-Moreira}}, \bibinfo{journal}{Phys. Rev. B}
  \textbf{\bibinfo{volume}{62}}, \bibinfo{pages}{115} (\bibinfo{year}{2000}).

\bibitem[{\citenamefont{Prozorov and Giannetta}(2006)}]{Prozorov2006}
\bibinfo{author}{\bibfnamefont{R.}~\bibnamefont{Prozorov}} \bibnamefont{and}
  \bibinfo{author}{\bibfnamefont{R.~W.} \bibnamefont{Giannetta}},
  \bibinfo{journal}{Supercond. Sci. Technol.} \textbf{\bibinfo{volume}{19}},
  \bibinfo{pages}{R41} (\bibinfo{year}{2006}).

\bibitem[{\citenamefont{Wang et~al.}(1966)\citenamefont{Wang, R, and
  Steinfink}}]{Wang}
\bibinfo{author}{\bibfnamefont{R.}~\bibnamefont{Wang}},
  \bibinfo{author}{\bibfnamefont{B.}~\bibnamefont{R}}, \bibnamefont{and}
  \bibinfo{author}{\bibfnamefont{H.}~\bibnamefont{Steinfink}},
  \bibinfo{journal}{Inorg. Chem.} \textbf{\bibinfo{volume}{5}},
  \bibinfo{pages}{1468} (\bibinfo{year}{1966}).

\bibitem[{\citenamefont{Whitehead}(1956)}]{HgMH}
\bibinfo{author}{\bibfnamefont{C.~S.} \bibnamefont{Whitehead}},
  \bibinfo{journal}{Proc. R. Soc. Lond. A} \textbf{\bibinfo{volume}{238}},
  \bibinfo{pages}{175} (\bibinfo{year}{1956}).

\bibitem[{\citenamefont{Budnick}(1960)}]{TaMH}
\bibinfo{author}{\bibfnamefont{J.~I.} \bibnamefont{Budnick}},
  \bibinfo{journal}{Phys. Rev.} \textbf{\bibinfo{volume}{119}}, \bibinfo{pages}{1578}
  (\bibinfo{year}{1960}).

\bibitem[{\citenamefont{Kozhevnikov et~al.}(2005)\citenamefont{Kozhevnikov,
  VanBael, Vinckx, Temst, VanHaesendonck, and Indekeu}}]{SnMH}
\bibinfo{author}{\bibfnamefont{V.~F.} \bibnamefont{Kozhevnikov}},
  \bibinfo{author}{\bibfnamefont{M.~J.} \bibnamefont{VanBael}},
  \bibinfo{author}{\bibfnamefont{W.}~\bibnamefont{Vinckx}},
  \bibinfo{author}{\bibfnamefont{K.}~\bibnamefont{Temst}},
  \bibinfo{author}{\bibfnamefont{C.} \bibnamefont{VanHaesendonck}},
  \bibnamefont{and} \bibinfo{author}{\bibfnamefont{J.~O.}
  \bibnamefont{Indekeu}}, \bibinfo{journal}{Phys. Rev. B}
  \textbf{\bibinfo{volume}{72}}, \bibinfo{pages}{174510}
  (\bibinfo{year}{2005}).

\bibitem[{\citenamefont{Hein and Raymond L.~Falge}(1961)}]{DPE}
\bibinfo{author}{\bibfnamefont{R.~A.} \bibnamefont{Hein}} \bibnamefont{and}
  \bibinfo{author}{\bibfnamefont{J.}~\bibnamefont{Raymond L.~Falge}},
  \bibinfo{journal}{Phys. Rev.} \textbf{\bibinfo{volume}{123}},
  \bibinfo{pages}{407} (\bibinfo{year}{1961}).

\bibitem[{\citenamefont{Tsujii et~al.}(2003)\citenamefont{Tsujii, Yoshimura,
  and Kosuge}}]{KW2}
\bibinfo{author}{\bibfnamefont{N.}~\bibnamefont{Tsujii}},
  \bibinfo{author}{\bibfnamefont{K.}~\bibnamefont{Yoshimura}},
  \bibnamefont{and} \bibinfo{author}{\bibfnamefont{K.}~\bibnamefont{Kosuge}},
  \bibinfo{journal}{J. Phys.: Condens. Matter} \textbf{\bibinfo{volume}{15}},
  \bibinfo{pages}{1993} (\bibinfo{year}{2003}).

\bibitem[{\citenamefont{McMillan}(1968)}]{McMillan}
\bibinfo{author}{\bibfnamefont{W.}~\bibnamefont{McMillan}},
  \bibinfo{journal}{Phys. Rev.} \textbf{\bibinfo{volume}{167}},
  \bibinfo{pages}{331} (\bibinfo{year}{1968}).

\bibitem[{\citenamefont{Cichorek et~al.}(2005)\citenamefont{Cichorek, Mota,
  Steglich, Frederick, Yuhasz, and Maple}}]{Cichorek2005}
\bibinfo{author}{\bibfnamefont{T.}~\bibnamefont{Cichorek}},
  \bibinfo{author}{\bibfnamefont{A.~C.} \bibnamefont{Mota}},
  \bibinfo{author}{\bibfnamefont{F.}~\bibnamefont{Steglich}},
  \bibinfo{author}{\bibfnamefont{N.~A.} \bibnamefont{Frederick}},
  \bibinfo{author}{\bibfnamefont{W.~M.} \bibnamefont{Yuhasz}},
  \bibnamefont{and} \bibinfo{author}{\bibfnamefont{M.~B.} \bibnamefont{Maple}},
  \bibinfo{journal}{Phys. Rev. Lett.} \textbf{\bibinfo{volume}{94}},
  \bibinfo{pages}{107002} (\bibinfo{year}{2005}).

\bibitem[{\citenamefont{Maple et~al.}(2007)\citenamefont{Maple, Henkie, Yuhasz,
  Ho, Yanagisawa, Sayles, Butch, Jeffries, and Pietraszko}}]{Maple2007}
\bibinfo{author}{\bibfnamefont{M.}~\bibnamefont{Maple}},
  \bibinfo{author}{\bibfnamefont{Z.}~\bibnamefont{Henkie}},
  \bibinfo{author}{\bibfnamefont{W.}~\bibnamefont{Yuhasz}},
  \bibinfo{author}{\bibfnamefont{P.-C.} \bibnamefont{Ho}},
  \bibinfo{author}{\bibfnamefont{T.}~\bibnamefont{Yanagisawa}},
  \bibinfo{author}{\bibfnamefont{T.}~\bibnamefont{Sayles}},
  \bibinfo{author}{\bibfnamefont{N.}~\bibnamefont{Butch}},
  \bibinfo{author}{\bibfnamefont{J.}~\bibnamefont{Jeffries}}, \bibnamefont{and}
  \bibinfo{author}{\bibfnamefont{A.}~\bibnamefont{Pietraszko}},
  \bibinfo{journal}{J. Magn. Magn. Mater.} \textbf{\bibinfo{volume}{310}},
  \bibinfo{pages}{182} (\bibinfo{year}{2007}).

\end{thebibliography}

\end{document}